\documentclass[twocolumn,showpacs,showkeys,preprintnumbers]{revtex4-1}
\usepackage{amssymb}

\usepackage{amsfonts}

\usepackage{latexsym}

\usepackage{dcolumn}

\usepackage{amsmath}

\usepackage{graphicx}

\usepackage{psfrag,pstricks}

\begin{document}

\title{Controlling steady-state bipartite entanglements and quadrature squeezing in a membrane-in-the-middle optomechanical system with two Bose-Einstein condensates}

\author{A. Dalafi$^{1}$ }
\email{adalafi@yahoo.co.uk}

\author{M. H. Naderi$^{2}$}

\affiliation{$^{1}$ Laser and Plasma Research Institute, Shahid Beheshti University, Tehran 1983969411, Iran\\
$^{2}$Quantum Optics Group, Department of Physics, Faculty of Science, University of Isfahan, Hezar Jerib, 81746-73441, Isfahan, Iran}

\date{\today}

\begin{abstract}
we study theoretically a driven hybrid optomechanical system with a membrane-in-the-middle configuration containing two identical elongated cigar-shaped Bose-Einstein condensates (BECs) in each side of the membrane. In the weakly interacting regime, the BECs can be considered as single-mode oscillators in the Bogoliubov approximation which are coupled to the optical field through the radiation pressure interaction so that they behave as two quasi-membranes. We show that the degree of squeezing of each BEC and its entanglement with the moving membrane can be controlled by the \textit{s}-wave scattering frequency of the other one. Since the \textit{s}-wave frequency of each BEC depends on the transverse trapping frequency of the atoms which is an experimentally controllable parameter, one can control the entanglement and squeezing of each BEC through the trapping frequency of the other one.

\end{abstract}


\maketitle
\section{Introduction}
%
%
There is an interesting correspondence between optomechanical systems, i.e., optical cavities with a moving end-mirror or with a membrane in the middle \cite{op1,op2,op3,op3,op4} from one hand and hybrid systems consisting of Bose-Einstein condensates (BECs) inside optical cavities from the other hand \cite{Bha 2009,Bha 2010}. In such hybrid systems, the excitation of a collective mode of the BEC couples to the radiation pressure of the cavity optical field \cite{Gupta,Brenn Nature, Kanamoto 2010}. Furthermore, such hybrid systems have provided a suitable background for the study of atom-photon interaction in the regime where their quantum mechanical properties are manifested in the same level \cite{Maschler2008, dom JOSA,Masch Ritch 2005}.

One of the advantages of cavities consisting of atomic ensembles in comparison to bare optomechanical systems is that in the former the atom-light interaction is enhanced because the atoms are collectively
coupled to the same optical mode \cite{dalafi1}. This collective mode which plays the role of the vibrational mode of a moving mirror or a membrane has the interesting capability that its coupling to the radiation pressure of the cavity can be increased by increasing the number of the atoms \cite{Biswas, Szirmai 2010, Nagy Ritsch 2009}.

On the other hand, different kinds of nonlinearities can be manifested in hybrid systems consisting of BEC \cite{Zhang 2009,dalafi7}. One of the most important nonlinear effects is due to the atom-atom interaction which can mostly affect the physical properties of the system \cite{dalafi3,dalafi4,dalafi5,dalafi6}.

In recent years, BEC-hybridized optomechanical setups, have emerged as an ideal platform for exploring the quantum phenomena at macroscopic level that are provided by the cooperation established between mechanical oscillators and atomic ensembles embedded into an optical cavity. Some important examples include ground-state cooling of the vibrational modes of a mechanical oscillator \cite{Mahajan PRA, Mahajan JPB,dalafi2, Yasir}, high fidelity quantum state transfer between a BEC and an optomechanical mirror \cite{Singh}, and quantum entanglement generation \cite{C1}. In light of the rich range of relevant physical effects emerging from such hybrid optomechanical systems, they are currently considered as basic building blocks for  quantum communication networks, quantum control and quantum state-engineering devices [for a recent review, see, for instance, Ref.\cite{Rogers}), and also as promising suitable candidates for investigating the foundations of quantum theory as well as testing its potential modifications \cite{Bassi}.

Motivated by the above-mentioned interesting features in the field of hybrid BEC-optomechanics, in this paper we are going to study a driven hybrid optomechanical system with a membrane-in-the-middle configuration containing two identical elongated cigar-shaped BECs in each side of the membrane. Since in the weakly interacting regime, the BECs can be considered as single-mode quantum oscillators in the Bogoliubov approximation which are coupled to the optical field through the radiation pressure interaction they behave as quasi-membranes.

In this way, the present system is equivalent to a bare optomechanical cavity containing three membranes inside, like the one studied in Ref.\cite{Seok}. However, the present hybrid system has the important advantage that the resonance frequency of each quasi-membrane depends on its \textit{s}-wave scattering frequency ($ \omega_{sw} $) of the nonlinear atom-atom interaction \cite{nagy2013} which is experimentally controllable by the transverse trapping frequency of the atoms \cite{Morsch}. We show that the degree of squeezing of each BEC and its entanglement with the moving membrane can be controlled by the \textit{s}-wave scattering frequency of the other one. 

\begin{figure}[ht]
\centering
\includegraphics[width=3.5in]{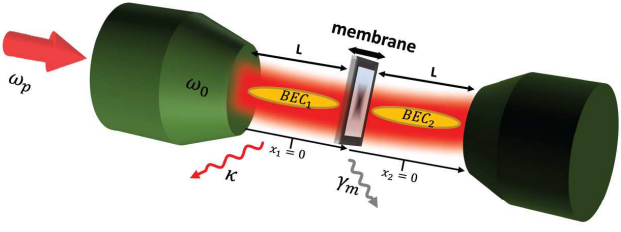} 
\caption{(Color online) Schematic of a membrane-in-the-middle  optomechanical cavity with length $2L$ containing two identical cigar-shaped BECs in each side of the membrane. The cavity which decays at rate $\kappa$ is driven through the left mirror by a laser with frequency $\omega_{p}$.}
\label{fig:fig1}
\end{figure}

The paper is structured as follows. In Sec. II we derive the Hamiltonian of the system. Then, in Sec. III the dynamics of the system is described and the quantum Langevin equations (QLEs) are derived and linearized around the semiclassical steady states. In Sec. IV we study the steady-state bipartite BEC-BEC and BEC-membrane entanglements as well as the quadrature squeezing properties of the two BECs. Finally, our conclusions are summarized in Sec. V.

\section{System Hamiltonian}

We consider a membrane-in-the-middle  optomechanical cavity with length $2L$ containing two identical cigar-shaped BECs in each side of the membrane, as schematically shown in Fig.\ref{fig:fig1}. The thin dielectric membrane with mass $ m $, frequency $ \omega_{m} $, and damping rate $ \gamma_{m} $ divides the cavity into two equal parts each with length $ L $. The two BECs on the left and the right side of the cavity each consisting of $ N_{j} $ two-level atoms ($ j=1,2 $ for the left  and right BEC, respectively) with mass $m_{a}$ and transition frequency $\omega_{a}$ are confined in cylindrical symmetric traps with transverse trapping frequencies $\omega_{\mathrm{\perp j}} (j=1,2)$ and negligible longitudinal confinement along the $x$ direction.

The cavity is driven at rate $\eta=\sqrt{2\mathcal{P}\kappa/\hbar\omega_{p}}$ through the left mirror by a laser with frequency $\omega_{p}$, and wavenumber $k=\omega_{p}/c$ ($\mathcal{P}$ is the laser power and $\kappa$ is the cavity decay rate) which excites a single mode of the cavity with natural frequency $ \omega_{0}=n\pi c/2L $ (in the absence of the middle membrane). It should be noted that for low values of the membrane reflectivity the cavity field can be considered as a single-mode field with central frequency $ \omega_{0}=n\pi c/2L $ \cite{Motazedifard}.  Moreover, in this type of optomechanical system, the frequency of the cavity field is dependent on the membrane displacement \cite{Jayich}. This dependence results in a nonlinear coupling, i.e., phonon number-dependent optomechanical coupling, between the cavity mode and the mechanical mode via multi-phonon excitations of the vibrational sidebands. However, in the limit of very small values of the Lamb-Dicke parameter $ \sigma=\frac{4\pi}{\lambda_{0}}\sqrt{\frac{\pi\hbar}{m\omega_{m}}} $ with $ \lambda_{0} $ being the wavelength of the cavity mode, and by considering the first excitation of the vibrational sideband the phonon-number dependence of the optomechanical coupling can be ignored \cite{Barzanjeh2011}.

In the dispersive regime where the laser pump is far detuned from the atomic resonance ($\Delta_{a}=\omega_{p}-\omega_{a}$  exceeds the atomic linewidth $\gamma$ by orders of magnitude), the excited electronic state of the atoms can be adiabatically eliminated and spontaneous emission can be neglected \cite{Masch Ritch 2004, Dom JB}. In this way, the dynamics of atoms can be described within an effective one-dimensional model by quantizing the atomic motional degree of freedom along the $x$ axis only. The Hamiltonian of the system can be written as
\begin{subequations}
\begin{eqnarray}
H&=&\hbar\omega_{0} a^{\dagger} a + i\hbar\eta(a e^{i\omega_{p}t}-a^{\dagger} e^{-i\omega_{p}t})\nonumber\\
&&+\hbar\omega_{m}b^{\dagger}b-\hbar\xi a^{\dagger}a(b+b^{\dagger})+H_{BEC}.\label{H1a}\\
H_{BEC}&=&\sum_{j=1}^{2}\int_{-L/2}^{L/2} dx_{j} \Psi^{\dagger}_{j}(x_{j})\Big[\frac{-\hbar^{2}}{2m_{a}}\frac{d^{2}}{dx^{2}_{j}}\label{H1b}\nonumber\\
&&+\hbar U_{0} \cos^2(kx_{j}) a^{\dagger} a+\frac{1}{2} U_{s}\Psi^{\dagger}_{j}(x_{j})\Psi_{j}(x_{j})\Big] \Psi_{j}(x_{j}).\nonumber\\
\end{eqnarray}
\end{subequations}

In Eq. (\ref{H1a}) the first term denotes the free Hamiltonian of the cavity mode, in which $ a (a^{\dagger}) $ is the photon annihilation (creation) operator $ ([a,a^{\dagger}]=1) $, the second term describes the pumping of the cavity by the external laser, the third term represents the free Hamiltonian of the oscillating membrane, with $ b (b^{\dagger}) $ being the annihilation (creation) operator of the mechanical mode ($ [b, b^{\dagger}]=1 $), and the fourth term describes the cavity-membrane interaction via radiation pressure with coupling rate of $ \xi $.

The last term in Eq.(\ref{H1a}) is the Hamiltonian of the two BECs which is given by Eq.(\ref{H1b}). The second quantized atomic wave fields $ \Psi_{j}(x_{j}) $ in Eq.(\ref{H1b}) with $ j=1,2 $ are, respectively, the annihilation operators of the first and the second BEC.  $U_{0}=g_{0}^{2}/\Delta_{a}$ is the optical lattice barrier height per photon which represents the atomic backaction on the field, $g_{0}$ is the vacuum Rabi frequency, $U_{s}=\frac{4\pi\hbar^{2} a_{s}}{m_{a}}$ and $a_{s}$ is the two-body \textit{s}-wave scattering length of atoms\cite{Masch Ritch 2004,Dom JB}. 

In the weakly interacting regime,where $U_{0}\langle a^{\dagger}a\rangle\leq 10\omega_{R}$ ($\omega_{R}=\frac{\hslash k^{2}}{2m_{a}}$ is the recoil frequency of the condensate atoms), and under the Bogoliubov approximation \cite{Nagy Ritsch 2009}, the atomic field operators of the BECs can be expanded as the following single-mode quantum fields
\begin{equation}\label{opaf}
\Psi_{j}(x_{j})=\sqrt{\frac{N_{j}}{L}}+\sqrt{\frac{2}{L}}\cos(2kx_{j}) c_{j},
\end{equation}
where the Bogoliubov mode $ c_{j} $ corresponds to the quantum fluctuations of the atomic field about the classical condensate mode ($ \sqrt{\frac{N_{j}}{L}} $). By substituting the atomic field operator of Eq.(\ref{opaf}) into Eq.(\ref{H1b}), the atomic part of Hamiltonian, i.e., $ H_{BEC} $ reduces to
\begin{eqnarray}\label{subH}
H_{BEC}&=&\sum_{j=1}^{2}\Big[\hbar\Omega_{j} c^{\dagger}_{j}c_{j}+\frac{\hbar}{\sqrt{2}}\zeta_{cj} a^{\dagger}a (c_{j}+c^{\dagger}_{j})\nonumber\\
&&+\frac{1}{4}\hbar\omega_{swj}(c^{2}_{j}+c^{\dagger 2}_{j})\Big],
\end{eqnarray}
where $ \Omega_{j}=4\omega_{R}+\omega_{swj} $ is the frequency of the Bogoliubov mode of the BEC on the side $ j $, $ \zeta_{cj}=\frac{1}{2}\sqrt{N_{j}}U_{0} $ is the optomechanical coupling between the Bogoliubov and the optical modes for either side of the membrane, and $ \omega_{swj}=8\pi\hbar a_{s}N_{j}/m_{a}Lw_{j}^2 $ (with $ w_{j} $ being the waist radius of the optical mode on the side $ j $) is the \textit{s}-wave scattering frequency of the atomic collisions on the either side of the membrane.

If we consider the Bogoliubov mode quadratures of the BECs as $Q_{cj}=(c_{j}+c^{\dagger}_{j})/\sqrt{2}$ and $P_{cj}=(c_{j}-c^{\dagger}_{j})/\sqrt{2}i$, then the Hamiltonian $ H_{BEC} $ can be written as
\begin{equation}\label{hc}
H_{BEC}=\sum_{j=1}^{2}\big[\frac{1}{2}\hbar\Omega_{j}^{(+)} Q_{cj}^{2}+\frac{1}{2}\hbar\Omega_{j}^{(-)} P_{cj}^{2}+\hbar\zeta_{cj} a^{\dagger}a Q_{cj}\big],
\end{equation}
where $ \Omega_{j}^{(\pm)}=\Omega_{j}\pm\frac{1}{2}\omega_{swj} $. Now, by defining new atomic quadratures as $ Q_{j}=\chi_{j} Q_{cj} $ and $ P_{j}=(1/\chi_{j})P_{cj} $ where $ \chi_{j}=\big(\frac{\Omega_{j}^{(+)}}{\Omega_{j}^{(-)}}\big)^{1/4} $, the Hamiltonian $ H_{BEC} $ gets the following form
\begin{equation}\label{Hc}
H_{BEC}=\sum_{j=1}^{2}\Big[\frac{1}{2}\hbar\omega_{j}(P_{j}^2+Q_{j}^{2})+\hbar\zeta_{j}a^{\dagger}a Q_{j}\Big].
\end{equation}
Based on Eq.(\ref{Hc}), the two BECs behave as two quantum harmonic oscillators with frequencies $ \omega_{j}=\sqrt{\Omega_{j}^{(+)}\Omega_{j}^{(-)}} $ which are coupled to the radiation pressure of the optical field with the optomechanical strength $ \zeta_{j}=\frac{1}{\chi_{j}}\zeta_{cj} $.

In the frame rotating with the pump frequency $ \omega_{p} $ the total Hamiltonian of the system, i.e., Eq.(\ref{H1a}) together with Eq.(\ref{Hc}) can be written as
\begin{eqnarray}\label{H}
H&=&\hbar\delta_{c} a^{\dagger} a + i\hbar\eta(a -a^{\dagger})+\hbar\omega_{m}b^{\dagger}b-\hbar\xi a^{\dagger}a (b+b^{\dagger})\nonumber\\
&&+\sum_{j=1}^{2}\Big[\frac{1}{2}\hbar\omega_{j}(P_{j}^2+Q_{j}^{2})+\hbar\zeta_{j}a^{\dagger}a Q_{j}\Big],
\end{eqnarray}
where $ \delta_{c}=-\Delta_{c}+\frac{1}{2}N U_{0} $ is the effective Stark-shifted detuning due to the presence of the BECs with $ N=N_{1}+N_{2} $ and $ \Delta_{c}=\omega_{p}-\omega_{0} $. The Hamiltonian of Eq.(\ref{H}) is similar to that of a three-mode optomechanical system consisting of three membranes inside an optical cavity where the Bogoliubov modes $ (Q_{j}, P_{j}) $ play the role of two quasi-membranes interacting with the radiation pressure of the cavity. In the atomic part of the Hamiltonian (\ref{H}), the nonlinear effect of atomic collisions has been coded in both the frequencies of the Bogoliubov modes of the two BECs, i.e.,  $ \omega_{j}=\sqrt{(4\omega_{R}+\frac{1}{2}\omega_{swj})(4\omega_{R}+\frac{3}{2}\omega_{swj})} $ and the optomechanical coupling constants $ \zeta_{j}=\frac{1}{\chi_{j}}\zeta_{cj} $ through the coefficients $ \chi_{j}=(\frac{4\omega_{R}+\frac{3}{2}\omega_{swj}}{4\omega_{R}+\frac{1}{2}\omega_{swj}})^{\frac{1}{4}} $.

\section{Dyanamics of The System}
The dynamics of the system described by the Hamiltonian in Eq. (\ref{H}) is fully characterized by the following set of nonlinear Heisenberg-Langevin equations:
\begin{subequations}\label{HaLaE}
\begin{eqnarray}
\dot{a}&=&-(i\delta_{c}+\kappa)a-\eta+i\xi a(b+b^{\dagger})\nonumber\\
&&-i\zeta_{1}a Q_{1}-i\zeta_{2}a Q_{2}+\sqrt{2\kappa}\delta a_{in},\label{NHLa}\\
\dot{b}&=&-(i\omega_{m}+\gamma_{m})b+i\xi a^{\dagger}a+\sqrt{2\gamma_{m}}\delta b_{in},\label{NHLb}\\
\dot{Q}_{j}&=&\omega_{j} P_{j}-\gamma_{c} Q_{j}+\sqrt{2\gamma_{c}} \delta Q_{j}^{in},\label{NHLc}\\
\dot{P}_{j}&=&-\omega_{j} Q_{j}-\gamma_{c} P_{j}-\zeta_{j} a^{\dagger}a+\sqrt{2\gamma_{c}} \delta P_{j}^{in},\label{NHLd}
\end{eqnarray}
\end{subequations}
where $ \gamma_{c} $ is  the dissipation rate of the collective density excitations of the BECs. The optical field quantum vacuum fluctuation $\delta a_{in}(t)$ satisfies the Markovian correlation functions, i.e., $\langle\delta a_{in}(t)\delta a_{in}^{\dagger}(t^{\prime})\rangle=(n_{ph}+1)\delta(t-t^{\prime})$, $\langle\delta a_{in}^{\dagger}(t)\delta a_{in}(t^{\prime})\rangle=n_{ph}\delta(t-t^{\prime})$ with the average thermal photon number $n_{ph}$ which is nearly zero at optical frequencies \cite{Gardiner}. Besides, $\delta b_{in}(t)$ is the quantum noise input for the moving membrane which also satisfies the same Markovian correlation functions as those of the optical noise \cite{K Zhang}. The input noise quadratures of the BECs are $ \delta Q_{j}^{in}=\frac{1}{\sqrt{2}}(\delta c_{j}^{in}+\delta c_{j}^{\dagger in}) $ and $ \delta P_{j}^{in}=\frac{1}{\sqrt{2}i}(\delta c_{j}^{in}-\delta c_{j}^{\dagger in}) $ where $ \delta c_{j}^{in} $ satisfies the same Markovian correlation functions as those of the optical and mechanical noises \cite{K Zhang, dalafi2}. The noise sources are assumed uncorrelated for the different modes of the BECs, mechanical and light fields. 

In order to linearize the nonlinear set of Eqs.(\ref{HaLaE}) we decompose each operator in Eqs. (\ref{NHLa}-\ref{NHLd}) as the sum of its steady-state value and a small fluctuation around its respective classical mean value. By substituting $ a=\alpha+\delta a $, $ b=\beta+\delta b $, $ Q_{j}=\bar{Q}_{j}+\delta Q_{j} $ and $ P_{j}=\bar{P}_{j}+\delta P_{j} $ into Eqs.(\ref{NHLa}-\ref{NHLd}) one can obtain a set of nonlinear algebraic equations for the
steady-state values,
\begin{subequations}\label{mv}
\begin{eqnarray}
\alpha&=&\frac{-\eta}{i\Delta+\kappa},\\
\beta&=&\frac{\xi |\alpha|^{2}}{\omega_{m}-i\gamma_{m}},\\
\bar{P}_{j}&=&\frac{\gamma_{c}}{\omega_{j}}\bar{Q}_{j},\\
\bar{Q}_{j}&=&-\frac{\gamma_{c}}{\omega_{j}}\bar{P}_{j}-\frac{\zeta_{j}}{\omega_{j}}|\alpha|^{2}.
\end{eqnarray}
\end{subequations}
Here, $ \Delta=\delta_{c}-2\xi\beta_{R}+\zeta_{1}\bar{Q}_{1}+\zeta_{2}\bar{Q}_{2} $ is the effective detuning of the cavity where $ \beta_{R} $ is the real part of $ \beta $. On the other hand, by defining the optical quadrature fluctuations as $ \delta X=\frac{1}{\sqrt{2}}(\delta a+\delta a^{\dagger}) $ and $ \delta Y=\frac{1}{\sqrt{2}i}(\delta a-\delta a^{\dagger}) $ and also the mechanical (membrane) quadrature fluctuations as $ \delta q=\frac{1}{\sqrt{2}}(\delta b+\delta b^{\dagger}) $ and $ \delta p=\frac{1}{\sqrt{2}i}(\delta b-\delta b^{\dagger}) $ one can obtain the linearized QLEs in the following compact matrix form
\begin{equation}\label{nA}
\delta\dot{u}(t)=A\delta u(t)+\delta n(t),
\end{equation}
where $ \delta u=[\delta X, \delta Y, \delta Q_{1}, \delta P_{1}, \delta Q_{2}, \delta P_{2}, \delta q, \delta p]^{T}  $ is the vector of continuous variable fluctuation operators and

\begin{widetext}
\begin{equation}
\delta n=[\sqrt{2\kappa}\delta X_{in}, \sqrt{2\kappa}\delta Y_{in}, \sqrt{2\gamma_{c}}\delta Q_{1}^{in}, \sqrt{2\gamma_{c}}\delta P_{1}^{in}, \sqrt{2\gamma_{c}}\delta Q_{2}^{in}, \sqrt{2\gamma_{c}}\delta P_{2}^{in}, \sqrt{2\gamma_{m}}\delta q_{in}, \sqrt{2\gamma_{m}}\delta p_{in}]^{T},
\end{equation}
is the corresponding vector of noises in which $\delta X_{in}=(\delta a_{in}+\delta a_{in}^{\dagger})/\sqrt{2}$ and $\delta Y_{in}=(\delta a_{in}-\delta a_{in}^{\dagger})/\sqrt{2}i$ are the input noise quadratures of the optical field and $\delta q_{in}=(\delta b_{in}+\delta b_{in}^{\dagger})/\sqrt{2}$ and $\delta p_{in}=(\delta b_{in}-\delta b_{in}^{\dagger})/\sqrt{2}i$ are the input noise quadratures of the mechanical mode of the membrane. The $ 8\times 8 $ matrix A is the drift matrix given by
\begin{equation}
A=\left(\begin{array}{cccccccc}
-\kappa & \Delta & \sqrt{2}\zeta_{1}\alpha_{I} & 0 & \sqrt{2}\zeta_{2}\alpha_{I} & 0 & -2\xi\alpha_{I} & 0 \\
-\Delta & -\kappa & -\sqrt{2}\zeta_{1}\alpha_{R} & 0 & -\sqrt{2}\zeta_{2}\alpha_{R} & 0 & 2\xi\alpha_{R} & 0 \\
0 & 0 & -\gamma_{c} & \omega_{1} & 0 & 0 & 0 & 0 \\
-\sqrt{2}\zeta_{1}\alpha_{R}& -\sqrt{2}\zeta_{1}\alpha_{I} & -\omega_{1} & -\gamma_{c} & 0 & 0 & 0 & 0\\
0 & 0 & 0 & 0 & -\gamma_{c} & \omega_{2} & 0 & 0\\
-\sqrt{2}\zeta_{2}\alpha_{R} & -\sqrt{2}\zeta_{2}\alpha_{I} & 0 & 0 & -\omega_{2} & -\gamma_{c} & 0 & 0\\
0 & 0 & 0 & 0 & 0 & 0 & -\gamma_{m} & \omega_{m}\\
2\xi\alpha_{R} & 2\xi\alpha_{I} & 0 & 0 & 0 & 0 & -\omega_{m} & -\gamma_{m}
  \end{array}\right),
\label{A}
\end{equation}
\end{widetext}
where $ \alpha_{R} $ and $ \alpha_{I} $ are, respectively, the real and imaginary parts of the optical mean field. The solutions to Eq.(\ref{nA}) are stable only if all the eigenvalues of the matrix $ A $ have negative real parts. The stability conditions can be obtained, for example, by using the Routh-Hurwitz criteria \cite{RH}.

\section{bipartite entanglements and squeezing}
In this section we first study how the Bogoliubov modes of the two BECs are entangled to each other and also we investigate the entanglement between each BEC and the moving membrane when the system reaches to its steady-state. Then, we examine the quadrature squeezing of the two BECs. For this purpose, one needs to obtain the correlation functions of the system in the stationary state in the regime where the system is stable. If all noises are assumed to be Gaussian, the linearized dynamics of the fluctuations leads to a zero-mean Gaussian steady state which is fully characterized by the $8\times8$ stationary correlation matrix (CM) $V$, with components $V_{ik}=\langle \delta u_i(\infty)\delta u_k(\infty)+\delta u_k(\infty)\delta u_i(\infty)\rangle/2 $. Using the QLEs, one can show that $ V $ fulfills the  Lyapunov equation \cite{Genes2008}
\begin{equation}\label{lyap}
AV+VA^T=-D,
\end{equation}
where
\begin{eqnarray}\label{D}
D&=&\mathrm{Diag}[\kappa,\kappa,\gamma_{c}(2n_{c1}+1),\gamma_{c}(2n_{c1}+1),\gamma_{c}(2n_{c2}+1),\nonumber\\
&&\gamma_{c}(2n_{c2}+1),\gamma_{m}(2n_{m}+1),\gamma_{m}(2n_{m}+1)],
\end{eqnarray}
is the diffusion matrix with $ n_{cj}=[\exp(\hbar\omega_{j}/k_{B}T-1)]^{-1} $ ($ j=1,2 $) as the mean number of thermal excitations of the Bogoliubov modes of the BECs and $ n_{m}=[\exp(\hbar\omega_{m}/k_{B}T-1)]^{-1} $ as the mean number of thermal phonons of the mechanical mode of the moving membrane. Equation(\ref{lyap}) is linear in $V$ and can straightforwardly be solved. However, the explicit form of $ V $ is complicated and is not reported here.

The bipartite entanglement between different degrees of freedom of the system can be calculated by using the logarithmic negativity \cite{eis}:
\begin{equation}\label{en}
E_N=\mathrm{max}[0,-\mathrm{ln} 2 \eta^-],
\end{equation}
where  $\eta^{-}\equiv2^{-1/2}\left[\Sigma(\mathcal{V}_{bp})-\sqrt{\Sigma(\mathcal{V}_{bp})^2-4 \mathrm{det} \mathcal{V}_{bp}}\right]^{1/2}$  is the lowest symplectic eigenvalue of the partial transpose of the $4\times4$ CM, $\mathcal{V}_{bp}$, associated with the selected bipartition, obtained by neglecting the rows and columns of the uninteresting mode,
\begin{equation}\label{bp}
\mathcal{V}_{bp}=\left(
     \begin{array}{cc}
     \mathcal{B}&\mathcal{C}\\
      \mathcal{C}^{T}&\mathcal{B}^{\prime}\\
       \end{array}
   \right),
\end{equation}
and $\Sigma(\mathcal{V}_{bp})=\mathrm{det} \mathcal{B}+\mathrm{det} \mathcal{B}^{\prime}-2\mathrm{det} \mathcal{C}$. In order to  calculate the bipartite entanglements, one should firstly solve for the set of Eqs.(\ref{mv}) to obtain the stationary mean values of the fields so that the matrix elements of the drift matrix $ A $ are determined. In this way, Eq.(\ref{lyap}) can be solved numerically.

Here, we analyze our results based on the experimentally feasible parameters given in Refs.\cite{Ritter Appl. Phys. B, Brenn Science},i.e., we assume each BEC consists of $ N_{i}=10^5 $ Rb atoms and the optical cavity has a length of $ 2L=374 \mu$m with bare frequency $ \omega_{0}=2.41494\times 10^{15} $Hz corresponding to a wavelength of $ \lambda_{0}=780 $nm and damping rate $ \kappa=2\pi\times 1.3 $MHz. The atomic $ D_{2} $ transition corresponding to the atomic transition frequency $ \omega_{a}=2.41419\times 10^{15} $Hz couples to the mentioned mode of the cavity. The atom-field coupling strength $ g_{0}=2\pi\times 14.1 $MHz and the recoil frequency of the atoms is $ \omega_{R}=23.7 $KHz. The moving membrane oscillates at frequency $ \omega_{m}=10^5 $Hz and has a damping rate of $ \gamma_{m}=2\pi\times 100 $Hz. We have also assumed that the temperature of the system is fixed at  $ T=0.1\mu $K.

\begin{figure}[ht]
\centering
\includegraphics[width=2.8in]{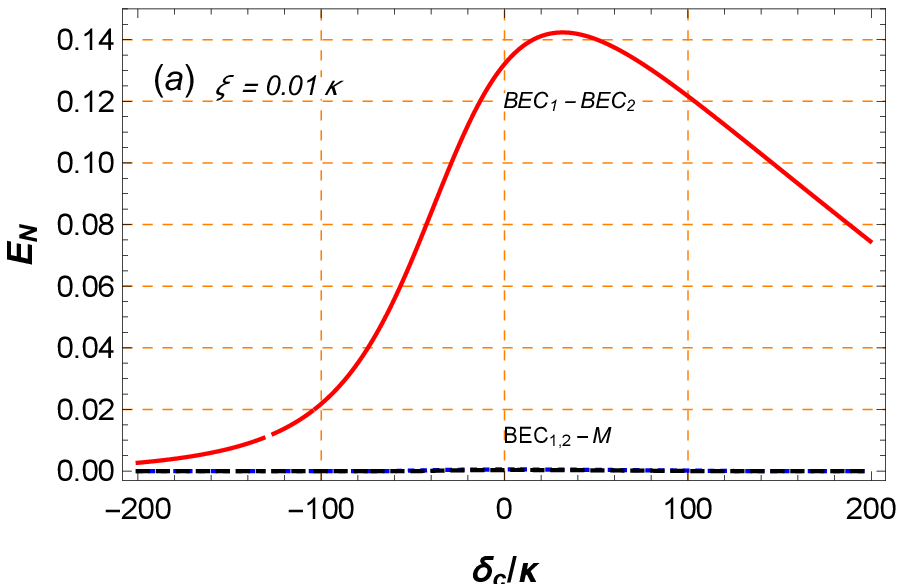}
\includegraphics[width=2.8in]{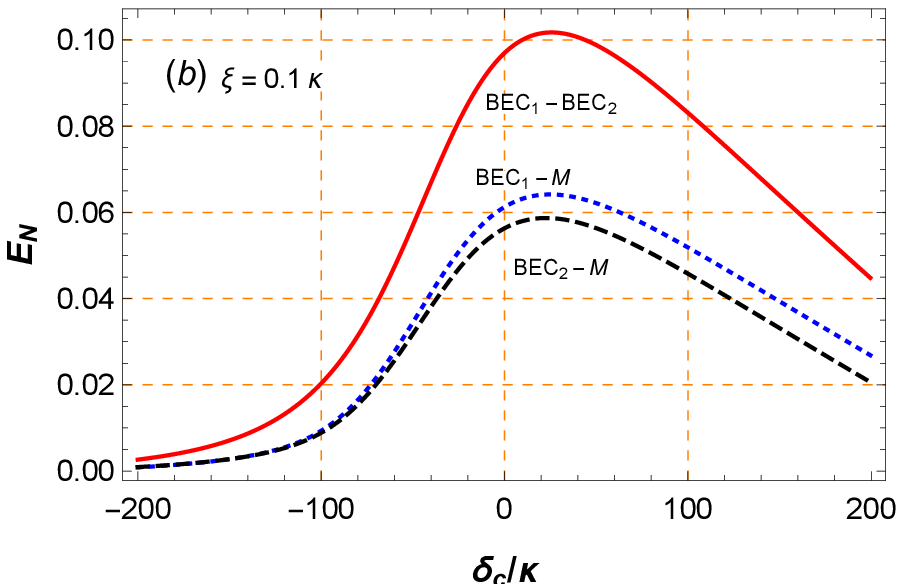}
\includegraphics[width=2.8in]{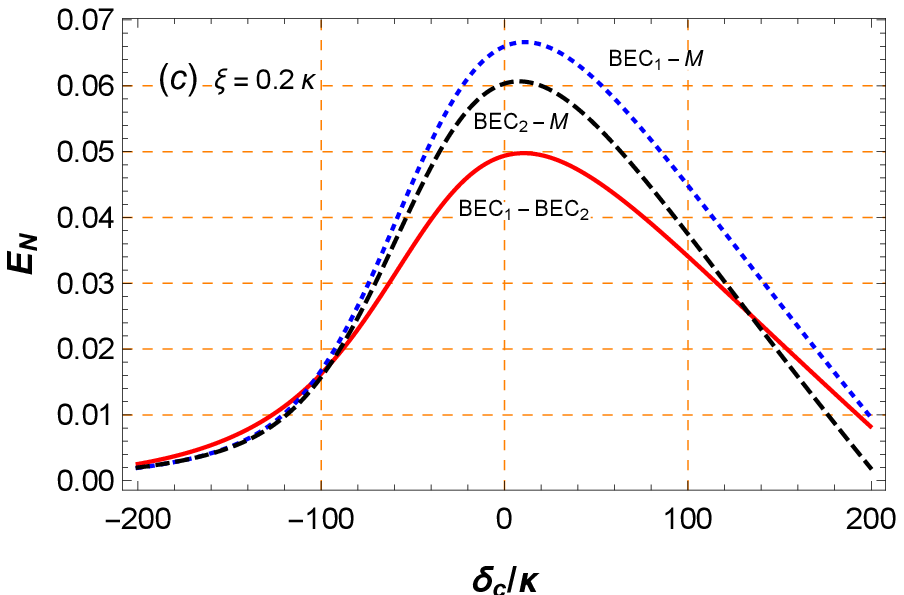}
\caption{
(Color online) The bipartite entanglement between the left and right BEC (red solid line), between the left BEC and the membrane (blue dotted line), between the right BEC and the membrane (black dashed line)  versus the normalized effective detuning $ \delta_{c}/\kappa $  for three different vaues of the optomechanical coupling: (a) $ \xi=0.01\kappa $, (b ) $ \xi=0.1\kappa $, and (c) $ \xi=0.2\kappa $ when $ \eta=100\kappa $, $ \omega_{sw1}=0.1\omega_{R} $ and $ \omega_{sw2}=0.2\omega_{R} $.}
\label{fig:fig2}
\end{figure}

In order to see how the optomechanical coupling between the membrane and the optical mode of the cavity affects the bipartite entanglements, in Fig.\ref{fig:fig2} we have plotted the bipartite entanglements between the two BECs and also between each BEC and the membrane for three different values of the optomechanical coupling when the cavity is pumped at rate $ \eta=100\kappa $ and the \textit{s}-wave scattering frequency of the first and the second BEC have been fixed, respectively, at $ \omega_{sw1}=0.1\omega_{R} $ and $ \omega_{sw2}=0.2\omega_{R} $. Here, the red solid line shows the entanglement between the first and the second BEC, the black dashed line shows the entanglement between the second BEC and the membrane, and the blue dotted line shows the entanglement between the first BEC and the membrane. The curves have been plotted versus the normalized cavity-pump detuning $ \delta_{c}/\kappa $ for three different values of the optomechanical coupling $ \xi=0.01\kappa $ [Fig.\ref{fig:fig2}(a)], $ \xi=0.1\kappa $ [Fig.\ref{fig:fig2}(b)], and $ \xi=0.2\kappa $ [Fig.\ref{fig:fig2}(c)].

As is seen from Fig.\ref{fig:fig2}(a) when the optomechanical coupling is weak ($ \xi=0.01\kappa $), there is no entanglement between the BECs and the membrane while there exists a strong entanglement between the two BECs in a wide range of the detuning which has a maximum value greater than 0.14 at $ \delta_{c}\approx 30\kappa $. However, by increasing the optomecanical coupling up to $ \xi=0.1\kappa $ [Fig.\ref{fig:fig2}(b)] the entanglement between the two BECs (red solid line) decreases a little bit while those of the BECs and the membrane (blue dotted line and black dashed line ) increase up to 0.06 near $ \delta_{c}\approx 30\kappa $. In Fig.\ref{fig:fig2}(c) where the optomechanical coupling is $ \xi=0.2\kappa $ the entanglements between the BECs and the membrane (blue dotted line and black dashed line ) grow higher than that of the two BECs (red solid line). 

Based on the results obtained in Fig.\ref{fig:fig2}, the weaker the optomechanical coupling between the optical field and the membrane, the stronger the entanglement between the two BECs. That is why the maximum entanglement between the two BECs is obtained when the optomechanical coupling between the membrane and the optical field is very weak. However, by increasing the optomechanical coupling the entanglement between the two BECs decreases while the entanglement between each BEC and the membrane increases. Furthermore, the \textit{s}-wave scattering frequency of the BECs can affect the amount of entanglement between each BEC and the membrane. In other words, with increasing the optomechanical coupling strength, the BEC with lower $ \omega_{sw} $ is more entangled with the membrane.

\begin{figure}[ht]
\centering
\includegraphics[width=2.6in]{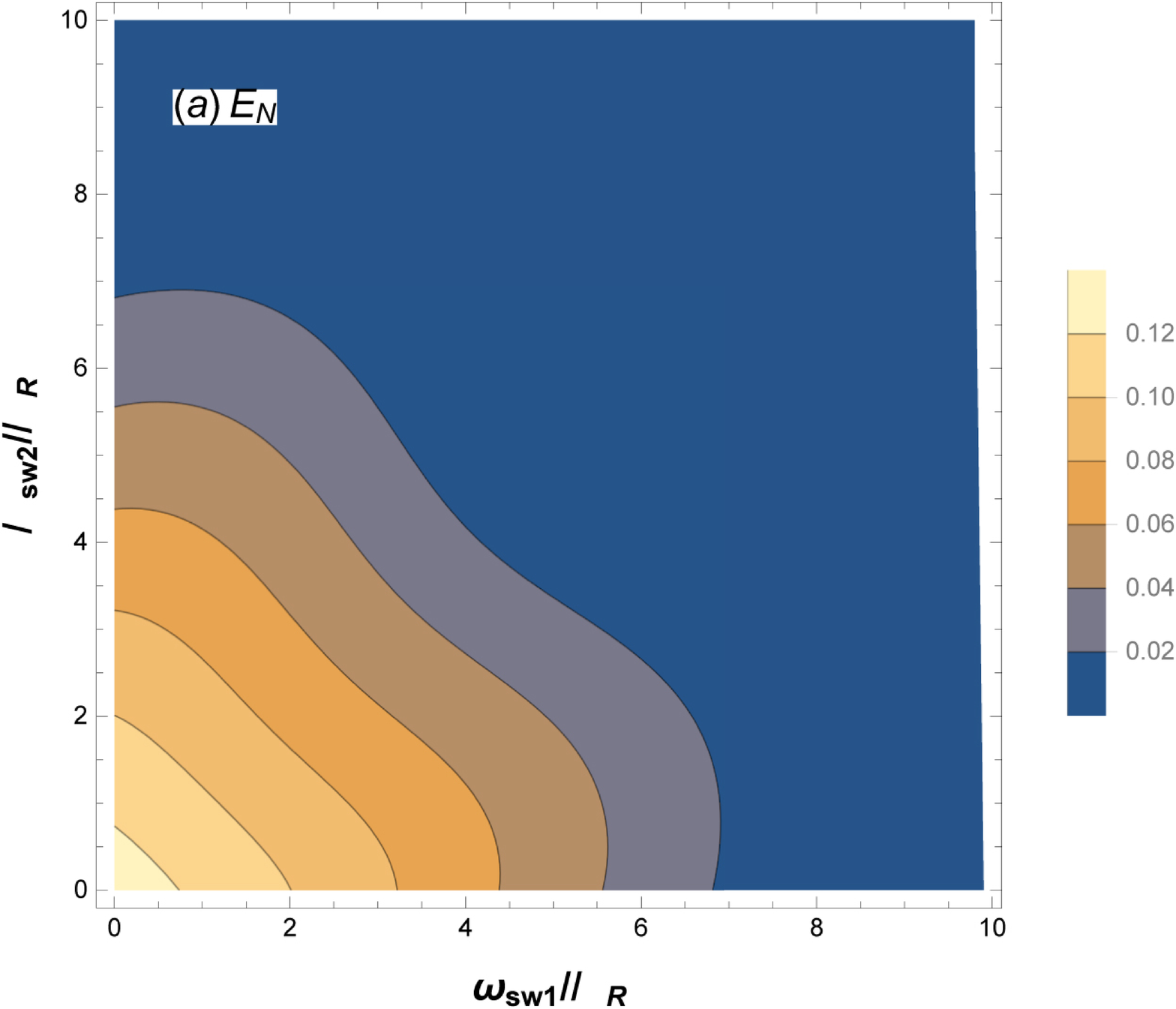}
\includegraphics[width=2.6in]{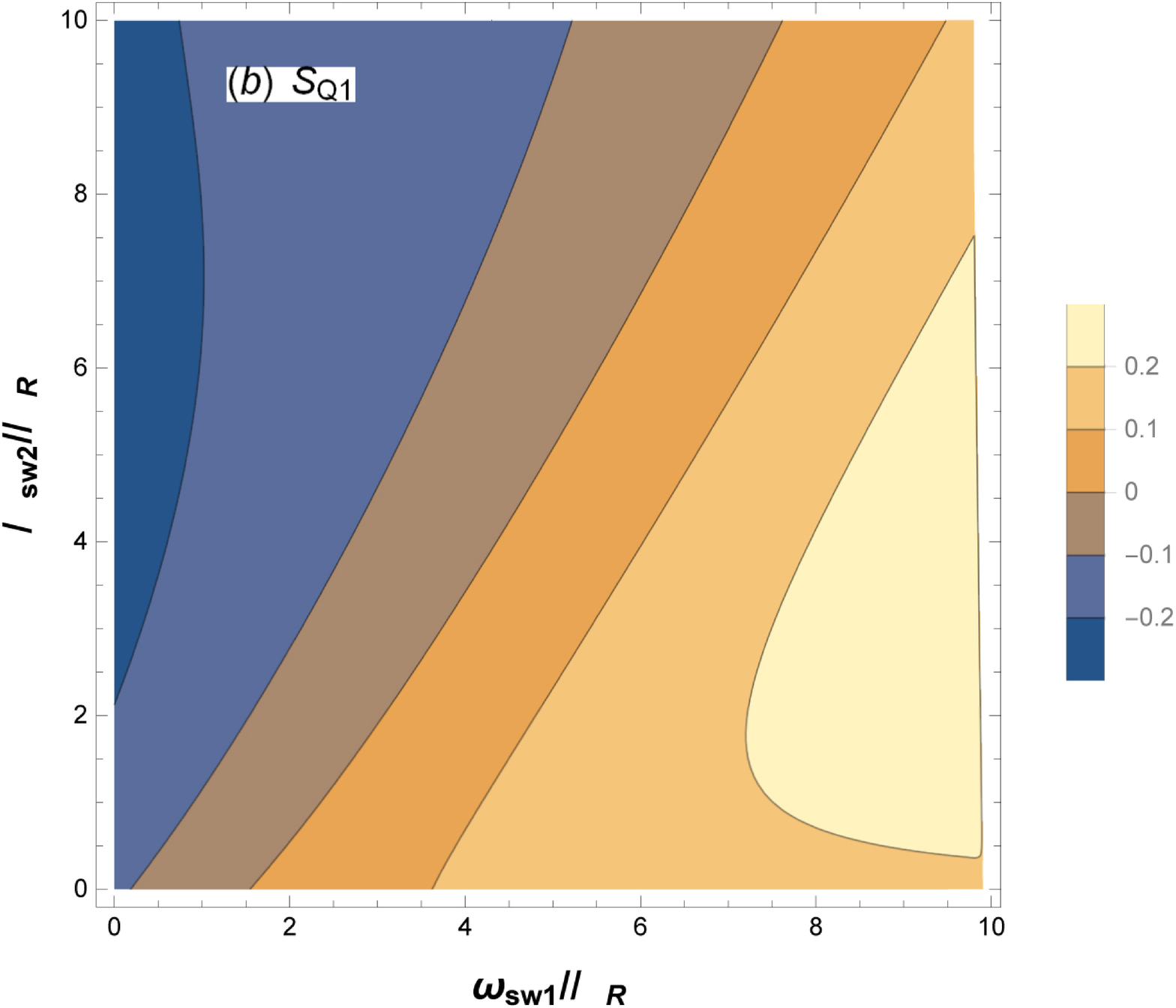}
\includegraphics[width=2.6in]{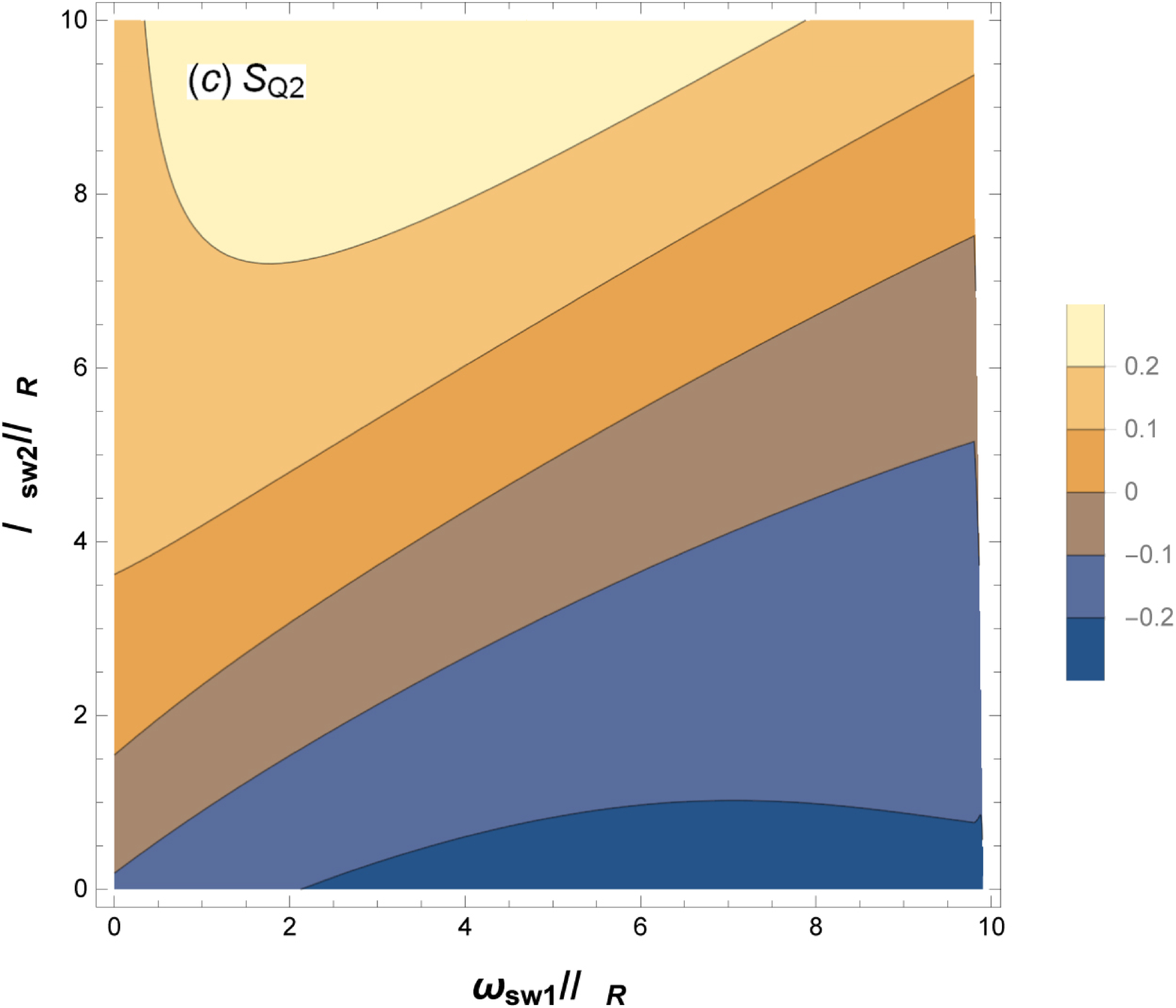}
\caption{
(Color online) (a) The contour plot of bipartite entanglement between the left and right BEC, (b) the squeezing parameter of the $ Q $ quadrature of the left BEC $ S_{Q1} $, and (c) the squeezing parameter of the $ Q $ quadrature of the right BEC $ S_{Q2} $ versus the normalized \textit{s}-wave scattering frequencies $ \omega_{sw1}/\omega_{R} $ and  $ \omega_{sw1}/\omega_{R} $ for $ \eta=100\kappa $, $ \delta_{c}=10\kappa $, $ \xi=0.05\kappa $. The other parameters are the same as those of Fig.\ref{fig:fig2}.}
\label{fig:fig3}
\end{figure}

Since the \textit{s}-wave scattering frequency of each BEC is a controllable parameter which can be adjusted experimentally by the transverse trapping frequency $\omega_{\mathrm{\perp j}}$ of that BEC \cite{Morsch}, it is interesting to study the variation of the BEC-BEC entanglement as a two-variable function of $ \omega_{sw1} $ and $ \omega_{sw2} $. For this purpose, in Fig.\ref{fig:fig3}(a) we have shown the bipartite entanglement between the two BECs as a contour plot versus the \textit{s}-wave scattering frequencies of the first ($ \omega_{sw1} $) and the second ($ \omega_{sw2} $) BEC. As is seen, the maximum amount of BEC-BEC entanglement is obtained for small values of the \textit{s}-wave scattering frequencies where $ \omega_{sw1}<0.5\omega_{R} $ and $ \omega_{sw2}<0.5\omega_{R} $. In this region the entanglement grows up to $ E_{N}\approx 0.12 $. However, by increasing $ \omega_{swj} $ the amount of entanglement decreases. The decrease in the BEC-BEC entanglement is because of the increase in the entanglement between one of the BECs and the membrane when the \textit{s}-wave scattering frequency of the other one increases.

Now, we explore the quadrature squeezing of the Bogoliubov modes of the two BECs. According to the Hamiltonian of Eq.(\ref{subH}), the atom-atom interaction in each BEC behaves as an atomic parametric amplifier in which the \textit{s}-wave scattering frequency plays the role of nonlinear gain parameter that can lead to the squeezing of the matter field of the BEC. As is seen from the total Hamiltonian of the system given in Eq.(\ref{H}), each BEC is a single-mode quantum field with quadratures $ Q_{j} $ and $ P_{j} $ obeying the commutation relations $ [Q_{j}, P_{k}]=i\delta_{jk} $. The degree of squeezing is defined in terms of the squeezing parameters $ S_{Qj}=2\langle(\Delta Q_{j})^{2}\rangle-1 $ and $ S_{Pj}=2\langle(\Delta P_{j})^{2}\rangle-1 $ where $ \langle(\Delta Q_{j})^{2}\rangle=\langle Q_{j}^2\rangle-\langle Q_{j}\rangle^2 $ and $ \langle(\Delta P_{j})^{2}\rangle=\langle P_{j}^2\rangle-\langle P_{j}\rangle^2 $ are the quantum uncertainties. Whenever $ S_{Qj}<0  $ or $ S_{Pj}<0  $, the corresponding quadrature is a squeezed one.

The squeezing parameters of the Bogoliubov modes of the two BECs can be expressed in terms of the stationary correlation matrix elements as follows
\begin{subequations}\label{SPQ}
\begin{eqnarray}
S_{Q_{1}}&=&2\langle\delta Q_{1}^{2}\rangle-1=2V_{33}-1,\\
S_{P_{1}}&=&2\langle\delta P_{1}^{2}\rangle-1=2V_{44}-1,\\
S_{Q_{2}}&=&2\langle\delta Q_{2}^{2}\rangle-1=2V_{55}-1,\\
S_{P_{2}}&=&2\langle\delta P_{2}^{2}\rangle-1=2V_{66}-1.
\end{eqnarray}
\end{subequations}

Our numerical results which are based on the experimental data of Refs.\cite{Brenn Science,Ritter Appl. Phys. B},show that there is no stationary squeezing in the quadratures $ \delta P_{j} $  while the squeezing occurs just for the quadratures $ \delta Q_{j} $. Therefore, in Figs.\ref{fig:fig3}(b) and (c) we have demonstrated, respectively, the squeezing parameters $ S_{Q1} $ and $ S_{Q2} $ as contour plots against $ \omega_{sw1} $ and $ \omega_{sw2} $. As is seen from Fig.\ref{fig:fig3}(b), the maximum degree of squeezing for the left BEC occurs in the region where $ \omega_{sw1}<\omega_{R} $ and $ \omega_{sw2}>2\omega_{R} $. In this region $ S_{Q1}<-0.2 $. On the other hand, the situation for the right BEC is vice versa, i.e., the maximum degree of squeezing occurs in the region where $ \omega_{sw1}>2\omega_{R} $ and $ \omega_{sw2}<\omega_{R} $. As is seen from Fig.\ref{fig:fig3}(c) in this region $ S_{Q2}<-0.2 $.

Based on these results, increasing the \textit{s}-wave scattering frequency in one of the BECs causes the degree of squeezing in the $ Q $ quadrature  of the other BEC to be increased if the \textit{s}-wave scattering frequency of the latter has been fixed at lower values. This phenomenon is a consequence of the quantum cross correlations between the two BECs that have been generated due to the interaction of the BECs with the optical field of the cavity. In this way, one can control the squeezing of each BEC through the \textit{s}-wave scattering frequency of the other one.

\begin{figure}[ht]
\centering
\includegraphics[width=2.8in]{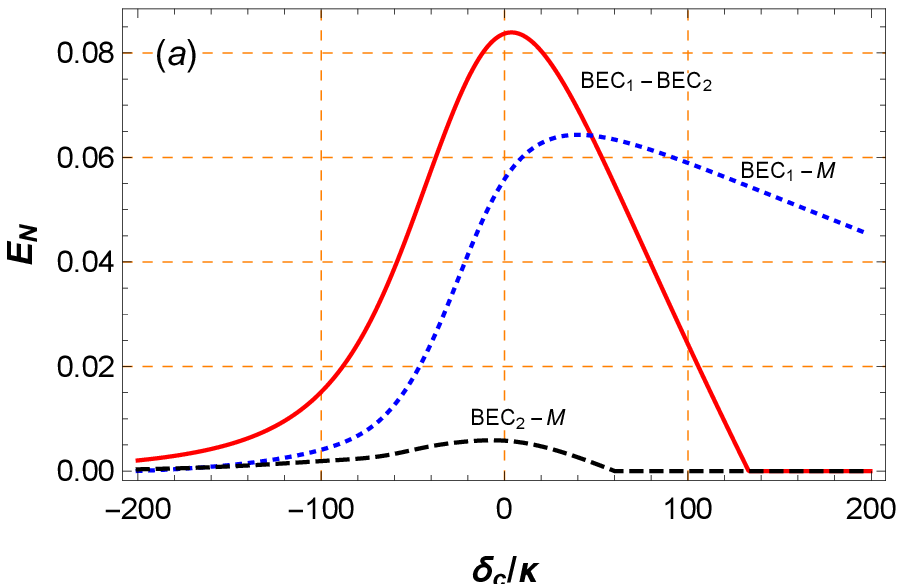}
\includegraphics[width=2.8in]{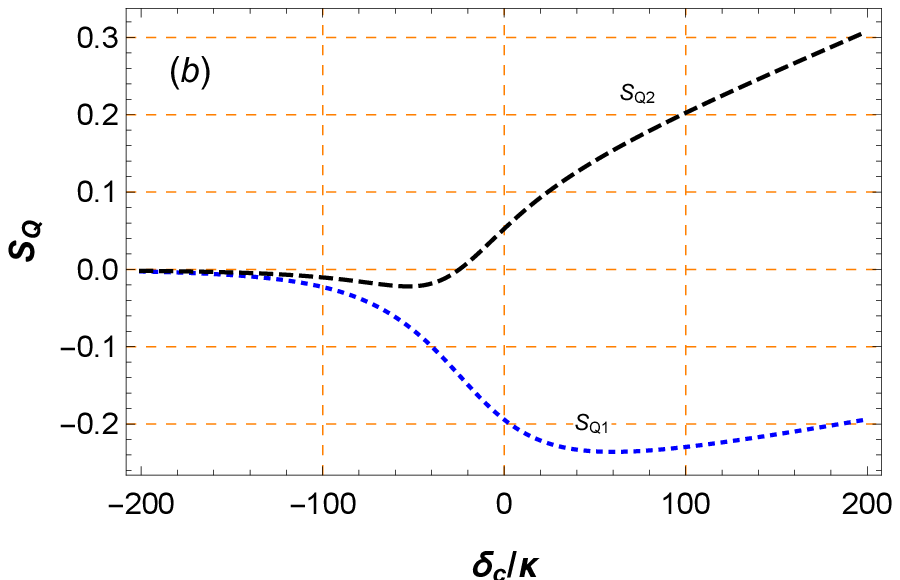}
\caption{
(Color online) (a) The bipartite entanglement between the left and right BEC (red solid line), between the left BEC and the membrane (blue dotted line), between the right BEC and the membrane (black dashed line), and (b) the squeezing parameters of the $ Q $ quadrature of the left ($ S_{Q1} $) and the right ($ S_{Q2} $) BEC versus the normalized effective detuning for two fixed values of $ \omega_{sw1}=0.1\omega_{R} $ and $ \omega_{sw2}=3\omega_{R} $. Here, we have assumed $ \xi=0.05\kappa $, $ \eta=100\kappa $ and the other parameters are the same as those of Fig.\ref{fig:fig2}}.
\label{fig:fig4}
\end{figure}

In order to investigate this matter more clearly, we have plotted in Figs.\ref{fig:fig4}(a) and \ref{fig:fig4}(b), respectively, the bipartite BEC-BEC as well as BEC-membrane entanglements and the squeezing parameters of the $ Q $ quadratures of the two BECs versus the normalized detuning $ \delta_{c}/\kappa $ for two fixed values of $ \omega_{sw1}=0.1\omega_{R} $ and $ \omega_{sw2}=3\omega_{R} $ when the cavity is pumped at rate $ \eta=100\kappa $ and the optomechanical coupling is $ \xi=0.05\kappa $. Here, at $ \delta_{c}=10\kappa $  the system parameters coincides with those of Fig.\ref{fig:fig3} in the region $ \omega_{sw1}<0.1\omega_{R}, \omega_{sw2}>2\omega_{R} $.

As is seen, when the \textit{s}-wave scattering frequency of the second BEC is higher than $ 2\omega_{R} $ (here $ \omega_{sw2}=3\omega_{R} $) while that of the first one has been fixed at the lower value $ \omega_{sw1}=0.1\omega_{R} $, the $ Q $ quadrature of the first BEC exhibits squeezing  ($ S_{Q1}<0 $) for a wide range of the detuning. Specifically, for $ 0 <\delta_{c}<180\kappa $ the squeezing parameter $ S_{Q1}<-0.2 $ [blue dotted line in  Fig.\ref{fig:fig4}(b)] . Instead, for the second BEC, having higher value of $ \omega_{sw} $, a small amount of $ Q $ quadrature squeezing occurs only in a limited range of detuning and it totally disappears ($ S_{Q2}>0 $) when $ \delta_{c}>-20\kappa $. More interestingly, the entanglement between the first BEC and the membrane [blue dotted line in  Fig.\ref{fig:fig4}(a)] goes to the maximum value where its $ Q $ quadrature has gained the maximum squeezing. The situation is in reverse for the second BEC whose entanglement with the membrane is decreased while its squeezing fades away.

As a physical interpretation of the above- mentioned phenomena, let us compare the present results with those obtained in Ref.\cite{dalafi6} where we have studied a more simplified setup consisting of an optomechanical cavity with a moving end mirror containing a single BEC. It was shown \cite{dalafi6} that in the regime similar to that considered here, the system behaves as an effective two-mode model in which the BEC and the mechanical mode are coupled to each other through the mediation of the optical field by an effective coupling parameter. 

Based on the results demonstrated in Fig.7(a) in Ref.\cite{dalafi6}, increasing the \textit{s}-wave scattering frequency makes the BEC-mirror entanglement be reduced. The reason was shown to be due to the dual effect of the atom-atom interaction: first, it strengthens the effective coupling parameter between the two modes, which should lead to an increase in the entanglement, and second, it makes the two modes get out of resonance. Since the latter effect dominates the former, the ultimate effect of the atom-atom interaction appears as a reduction of the BEC-mirror entanglement. 

A similar phenomenon takes place in the present setup when the \textit{s}-wave scattering frequency of one of the BECs is higher than the other one's. Because the BEC with higher $ \omega_{sw} $ (the second BEC in Fig.\ref{fig:fig4}), is more out of resonance with the membrane, it has a lower entanglement compared to the other one. On the other hand, the increase in the entanglement between the first BEC (with lower $ \omega_{sw} $) and the membrane, leads to the reduction in the BEC-BEC entanglement. That is why the entanglement between the two BECs reduces in Fig.\ref{fig:fig3}(a) when $ \omega_{sw1} $ or $ \omega_{sw2} $ is increased.

In short, the degree of quadrature squeezing of each BEC and its entanglement with the moving membrane can be controlled by the \textit{s}-wave scattering frequency of the other one. Since the \textit{s}-wave frequency of each BEC depends on the transverse trapping frequency of the atoms which is an experimentally controllable parameter, one can control the entanglement and squeezing of each BEC through the trapping frequency of the other one.

\section{Conclusions}
In conclusion, we have studied a driven hybrid optomechanical setup with a membrane-in-the-middle containing two identical cigar-shaped BECs in each side of the membrane. In the weakly interacting regime, each BEC can be considered as a single-mode oscillator in the Bogoliubov approximation. In this way, the Bogoliubov mode of each BEC is coupled to the optical field through a radiation pressure term and behaves as a quasi-membrane. 

We have shown that the degree of quadrature squeezing of each BEC and its entanglement with the moving membrane can be controlled by the \textit{s}-wave scattering frequency of the other one. If the \textit{s}-wave scattering frequency of one BEC is increased its degree of quadrature squeezing and its entanglement with the membrane is decreased. Instead, the other BEC with lower \textit{s}-wave scattering frequency will have higher degree of quadrature squeezing and also higher entanglement with the membrane. Furthermore, with increasing the \textit{s}-wave scattering frequency of one or both BECs, the entanglement between the two BECs is reduced.

Since the \textit{s}-wave scattering frequency of each BEC depends on the transverse trapping frequency of the atoms which is an experimentally controllable parameter, one can control the entanglement and squeezing of each BEC through the trapping frequency of the other one.

\section*{Acknowledgement}
A.D wishes to thank the Laser and Plasma Research Institute of Shahid Beheshti University for its support.

\bibliographystyle{apsrev4-1}

\end{document}